\documentclass[aps,prd,onecolumn,showpacs,groupedaddress]{revtex4-2}
\usepackage{graphicx}
\usepackage{dcolumn}
\usepackage{bm}
\usepackage{color}
\usepackage{longtable}
\usepackage{supertabular}
\usepackage[T1]{fontenc}
\usepackage{epstopdf}
\usepackage{amsmath}
\usepackage{amssymb}
\usepackage{setspace}
\usepackage{multirow}
\usepackage{booktabs}
\usepackage[section]{placeins}
\usepackage{mathrsfs}
\usepackage{hyperref}
\usepackage{mnsymbol}

\makeatletter

\newcommand{\Rmnum}[1]{\expandafter\@slowromancap\romannumeral #1@} 
\newcommand{\bq}{\begin{equation}}
\newcommand{\eq}{\end{equation}}
\newcommand{\bqn}{\begin{eqnarray}}
\newcommand{\eqn}{\end{eqnarray}}

\makeatother

\begin{document}

\title{Suppression of the multiplicity fluctuations in particle correlations}

\author{Chong Ye$^{1,2}$}
\author{Hong-Hao Ma$^{3,4}$}
\author{Dan Wen$^{5}$}
\author{Philipe Mota$^{6}$}
\author{Wei-Liang Qian$^{7,2,1}$}\email[E-mail: ]{wlqian@usp.br}
\author{Rui-Hong Yue$^{1}$}

\affiliation{$^{1}$ Center for Gravitation and Cosmology, College of Physical Science and Technology, Yangzhou University, Yangzhou 225009, China}
\affiliation{$^{2}$ Faculdade de Engenharia de Guaratinguet\'a, Universidade Estadual Paulista, 12516-410, Guaratinguet\'a, SP, Brazil}
\affiliation{$^{3}$ Department of Physics, Guangxi Normal University, Guilin 541004, China}
\affiliation{$^{4}$ Guangxi Key Laboratory of Nuclear Physics and Technology, Guangxi Normal University, Guilin 541004, China}
\affiliation{$^{5}$ College of Science, Chongqing University of Posts and Telecommunications, 400065, Chongqing, China}
\affiliation{$^{6}$ Instituto de F\'isica, Universidade Federal do Rio de Janeiro, C.P. 68528, 21945-970, Rio de Janeiro-RJ , Brazil}
\affiliation{$^{7}$ Escola de Engenharia de Lorena, Universidade de S\~ao Paulo, 12602-810, Lorena, SP, Brazil}

\begin{abstract}
Multiplicity fluctuations play a crucial role in relativistic heavy-ion collisions.
In this work, we explore how the multiplicity fluctuations can be effectively suppressed in the measurement of particle correlations.
In particular, through proper normalization, particle correlations can be evaluated in a manner irrelevant to multiplicity.
When the multiplicity fluctuations are adequately extracted, Monte Carlo simulations show that the remaining correlations possess distinct features buried in the otherwise overwhelming fluctuations.
Moreover, we argue that such a normalization scheme naturally agrees with the multi-particle correlator, which can be evaluated using the Q-vectors.
The implications of the present study in the data analysis are also addressed.
\end{abstract}

\date{March 16th, 2023}

\maketitle
\newpage

\section{Introduction}\label{section1}

Collectivity is one of the most prominent features of the hot and dense matter produced in relativistic heavy-ion collisions.
It is characterized by the flow harmonics that can be extracted from the hadrons' azimuthal distribution and correlations.
It is understood that such collective flow is primarily generated during the hydrodynamic or transport evolution stage, where the dynamics of the strongly coupled quark-gluon plasma is governed by an effective description attained at the long wavelength limit~\cite{hydro-review-04, hydro-review-05, hydro-review-06, sph-review-01, sph-review-02, hydro-review-07, hydro-review-08, hydro-review-09, hydro-review-10}.
As an effective theory, the system's degrees of freedom are the components of the continuum's energy-momentum tensor.
The relevant system of equations of motion is closed by further furnishing the information and assumptions regarding the matter's equation of state and viscosities. In principle this can be derived from the underlying microscopic theory.
Hydrodynamical evolution plays a crucial role in understanding the relationship between the empirical data from the heavy-ion collisions and the system's initial conditions. 
As the input of the hydrodynamic model, the initial conditions are expressed mainly in terms of the energy-momentum tensor.
The hydrodynamic evolution is responsible for transforming the initial-state geometric deformation and fluctuations into the final-state anisotropy in the momentum space. 
In this regard, it is a consensus that the collective flow and particle correlations carry crucial information on the hot and dense system created in the heavy-ion collisions~\cite{hydro-vn-07, hydro-vn-08, hydro-v3-01, hydro-v3-02, sph-eos-02, sph-cfo-01, sph-corr-ev-10, sph-corr-03, sph-corr-04, sph-corr-05, sph-corr-ev-06, sph-corr-ev-08, sph-corr-09}.

The collective flow is defined in terms of flow harmonics $v_n$ which are the Fourier coefficients of the one-particle distribution function in azimuthal angle $\varphi$~\cite{event-plane-method-1}
\begin{eqnarray}
\frac{dN}{d\varphi}(\varphi)=\frac{N}{2\pi}\left[1+\sum_{n=1}2v_{n}\cos{n(\varphi-\Psi_n)}\right] ,
\label{oneParDis}
\end{eqnarray}
where $N$ is the multiplicity of a given event, and the event planes $\Psi_n$ are generally different for individual harmonic orders. 
Different harmonic coefficients carry different physical interpretations.
The elliptic flow $v_2$ is due mainly to the initial geometric shape of the collision system~\cite{hydro-vn-08, hydro-v2-shuryak-01, hydro-v2-voloshin-01}.
The latter is speculated to have an almond shape dictated by the overlap area of the two nuclei.  
On the other hand, the triangular flow $v_3$ is primarily attributed to the fluctuations of the initial conditions~\cite{hydro-v3-01}, taking place on an event-by-event basis.
Various numerical methodologies have been developed to deduce the value of flow harmonics $v_{n}$ from the experimental data. 
The classic event plane method~\cite{event-plane-method-1, Poskanzer:1998yz} is structured to approximate the event planes $\Psi_n$ as represented in Eq.\eqref{oneParDis}. 
A salient feature of this method is tied to the inherent challenge that the reaction plane~\cite{Alver:2010gr} is not directly observable in experiments.
Furthermore, a few methods involve the concepts of Q-vectors and cumulants~\cite{hydro-vn-07, Borghini:2000sa, Bilandzic:2010jr, Jia:2017hbm}. 
A notable advantage of these approaches is their capability to obviate the event planes from Eq.\eqref{oneParDis}. 
Also, the expression for the cumulant can be succinctly portrayed through the generating function~\cite{Borghini:2000sa}. 
The relevant methods consist of particle cumulants~\cite{Borghini:2000sa, Bilandzic:2010jr}, Lee--Yang zeros~\cite{Bhalerao:2003xf, Bhalerao:2003yq, FOPI:2005ukb}, symmetric cumulants~\cite{Bilandzic:2013kga}, as well as their generalizations~\cite{Bhalerao:2013ina, DiFrancesco:2016srj, Mordasini:2019hut}.
Maximum likelihood estimation can also be utilized to analyze the flow harmonics~\cite{sph-vn-10}.
Extensive efforts, including numerical simulations, have been carried out to investigate the collective flow. These include the elliptic~\cite{hydro-v2-heinz-01, hydro-v2-heinz-05, hydro-v2-hirano-01, hydro-v2-hirano-05}, triangular~\cite{hydro-v3-03, hydro-v3-04, hydro-v3-05, hydro-v3-06, hydro-v3-08}, and higher order harmonic coefficients~\cite{hydro-vn-32, hydro-vn-33, hydro-vn-34, hydro-vn-35, hydro-vn-36, hydro-vn-38, hydro-vn-43, hydro-vn-44}.
The connection between the initial geometric anisotropy and the final-state flow harmonics has been a relevant topic in the literature~\cite{hydro-v3-02, hydro-vn-34, hydro-vn-42, hydro-vn-45, sph-corr-ev-04, sph-vn-04, hydro-corr-04, hydro-corr-05, sph-corr-ev-06, sph-corr-ev-08, sph-corr-09}.

A pertinent feature of the collective flow resides in its fluctuations. 
On the one hand, these fluctuations carry crucial information on the initial eccentricities associated with the dynamics of the partonic degrees of freedom governed by the underlying microscopic model.
On the other hand, even though one has an infinite number of events, for an individual event finite multiplicity also gives rise to additional fluctuations~\cite{sph-vn-09, sph-vn-10} owing to statistical uncertainty.
Specifically, as we average over a significant number of events, the variance of the flow harmonics remains finite.
The latter does not vanish even when the event-by-event fluctuations in the initial conditions are entirely eliminated.
Even though these two types of fluctuations are distinct in their physical origin they might be indistinguishable from an empirical point of view. 

Particle correlation at intermediate and low transverse momentum is another relevant observable that provides valuable information on the properties of the created medium.
In practice, the correlated di-hadron yields are expressed in the pseudo-rapidity difference $\Delta\eta$ and azimuthal angular spacing $\Delta\varphi$. 
Strong enhancement was observed~\cite{RHIC-star-ridge-01,RHIC-star-ridge-02,RHIC-star-ridge-03,RHIC-phenix-ridge-04,RHIC-phenix-ridge-05,RHIC-phobos-ridge-06,RHIC-phobos-ridge-07,LHC-alice-vn-04,LHC-cms-ridge-05,LHC-atlas-vn-01} compared to those at high transverse momentum~\cite{RHIC-star-jet-01,RHIC-star-jet-02}. 
The correlation on the near side is usually referred to as ``ridge'', a manifestation of long-range correlation in pseudo-rapidity.
The structure features a narrow $\Delta\varphi$ peak and a sizable extension in $\Delta\eta$ about a few units.  
On the away side the observed correlations present interesting features, usually known as ``shoulders''.
Specifically, a double peak is observed for certain centralities and transverse-momentum ranges.
These structures, for the most part, can be interpreted as a medium effect owing to the hydrodynamical evolution of the system and in terms of the triangular flow~\cite{sph-corr-01,sph-corr-03,hydro-v3-01,hydro-v3-02,hydro-v3-03,hydro-v3-04,hydro-v3-05,sph-corr-04,sph-corr-05,sph-corr-06,hydro-v3-06,hydro-v3-08,sph-corr-09}. 
Besides the A+A collisions discussed above, the long-range correlation and collective flow have also been observed in small systems associated with p+A and p+p collisions by the CMS~\cite{LHC-cms-ridge-01, LHC-cms-ridge-07, LHC-cms-ridge-08, LHC-cms-ridge-09}, ATLAS~\cite{LHC-atlas-ridge-01, LHC-atlas-ridge-02}, and ALICE~\cite{LHC-alice-ridge-01, LHC-alice-ridge-02, LHC-alice-ridge-05}.
It is important to note that the observed ridge is more pronounced in high-multiplicity events.
Efforts have been made to understand the main features of the data in terms of gluon saturation~\cite{Dusling:2012cg, hydro-vn-38}, parton-induced interactions~\cite{Arbuzov:2011yr}, string shoving~\cite{Bierlich:2016vgw, Bierlich:2017vhg}, collective phenomena due to hydrodynamic~\cite{Weller:2017tsr, Zhao:2017rgg, Mantysaari:2017cni} or transport~\cite{Greif:2017bnr} behavior, as well as by simulations such as PYTHIA8~\cite{pythia-5}.

Specific features of the long-range correlations have been explored more thoroughly by investigating their centrality and trigger-angle dependence. 
PHENIX collaboration analyzed the resulting two-particle correlations as a function of centrality~\cite{RHIC-phenix-ridge-05}, defined in terms of multiplicity.
As the system goes from the central to more peripheral collisions, the away side of the two-particle correlations evolves from one single peak to a double peak.
STAR Collaboration reported measurements~\cite{RHIC-star-plane-01} of two-particle azimuthal correlations as a function of the trigger particle's azimuthal angle concerning the event plane, denoted by $\varphi_s=|\varphi_\mathrm{tr}-\Psi_\mathrm{EP}|$, evaluated for different trigger and associated transverse momenta $p_T$. 
Later, the analysis~\cite{RHIC-star-plane-03, RHIC-star-plane-02} was developed further with an attempt to separate the ``jet'' from the ``ridge''.
In other words, the motivation of the study is to obtain the resulting two-particle correlations due to the mini-jets by removing the contributions owing to the collective flow.
Specifically, the ridge yields are extracted by introducing a rapidity gap $|\Delta\eta|$ while counting hadron pairs. While implementing the procedure, one also assumes that the ridge possesses a uniform in $\Delta\eta$ direction, but the jet does not. 
In Refs.~\cite{RHIC-star-plane-03, RHIC-star-plane-02}, a cut of $\Delta\eta > 0.7$ is applied on the pseudorapidity difference between the trigger and associated particles aiming at minimizing the near-side jet contributions.
The above recipe is qualitatively in accordance with the theoretical estimations~\cite{Salgado:2003rv, Vitev:2008rz, KunnawalkamElayavalli:2017hxo, Brewer:2018mpk, Chang:2019sae, Luo:2021hoo} and experimental measurements~\cite{LHC-cms-jet-10, LHC-cms-jet-15, RHIC-star-jet-15, RHIC-star-jet-16, Zardoshti:2019cvh} on jet shapes, particularly the event-plane dependence of jet shapes explored by STAR collaboration~\cite{RHIC-star-jet-16}.
Nonetheless, in applying such a refinement the residual correlations exhibit some intriguing characteristics.
The most recent study (c.f. Fig.~7 of~\cite{RHIC-star-plane-02}) showed that the away-side structure evolves from the in-plane to out-of-plane directions.
To be more specific, the correlations vary with $\varphi_s$ on both the near side and away side of the trigger particle. 
The near-side peak drops when the trigger particle goes from in-plane to out-of-plane, and the correlations in the away-side might evolve from single- to double-peak with increasing $\varphi_s$. 

Interestingly, hydrodynamic calculations using the SPheRIO code~\cite{sph-cfo-01,sph-v2-01, sph-v2-02, sph-v1-01, sph-vn-01,sph-v2-03,sph-hbt-01} provides a good description~\cite{sph-corr-01,sph-corr-02,sph-corr-03,sph-corr-04,sph-corr-05,sph-corr-06,sph-corr-08,sph-corr-09,sph-corr-ev-04,sph-corr-ev-05,sph-corr-ev-06,sph-corr-ev-08} for many observed characteristics of the ridge in A+A collisions.
Besides reproducing the results numerically using the hydrodynamic approach~\cite{sph-corr-01}, many features related to the long-range correlations can be explained in terms of a peripheral tube model~\cite{sph-corr-03,sph-corr-04}.
Analytic calculations have shown that correlations of different origins can be primarily viewed to contribute in a separate fashion, giving rise to the observed ridge and shoulder.
The model was further elaborated to discuss the event-plane dependence~\cite{sph-corr-ev-04} and centrality dependence~\cite{sph-corr-ev-06, sph-corr-ev-08} where the rapidity gas was also considered.
Regarding the peripheral tube model, the evolution of the correlation structure on the away side is attributed to an interplay between triangular flow and multiplicity fluctuations.
As discussed in~\cite{sph-corr-ev-04, sph-corr-ev-06}, the resultant two-particle correlations consist of two contributions.
The multiplicity fluctuations of the background dominate the first one, while the second can be attributed to the peripheral tube.
However, the magnitude of the second term does not affect the overall multiplicity.
Such an idea has been confirmed and further explored in terms of realistic simulations~\cite{sph-corr-08, sph-corr-09}.
Moreover, the approach based on the peripheral tube model agrees very well with the qualitative properties of the Fourier coefficients of the two-particle correlations extracted from the data~\cite{hydro-corr-ph-21}.
These results indicate that the contributions associated with the overall multiplicity fluctuations separate from the remaining contribution related to the eccentricities in the initial conditions.
The information carried by the latter can be extracted if the multiplicity fluctuations are eliminated. 
The present study is mainly motivated to elaborate a scheme to suppress the multiplicity functions from the two-particle correlations.
As a result, one gains direct access to particle correlations associated with the initial geometric fluctuations.

The remainder of the present paper is organized as follows. 
The following section briefly reviews the peripheral tube model and its role in interpreting the observed features in the two-particle correlations.
In Sec.~\ref{section3}, we elaborate on a scheme of normalization in calculating particle correlations to eliminate the multiplicity fluctuations.
Also, we discuss the normalized Q-vectors.
Numerical simulations are carried out and demonstrate how the scheme works.
The last section is devoted to further discussions and concluding remarks.  

\section{The peripheral tube model}\label{section2}

The peripheral tube model~\cite{sph-review-02} provides an intuitive picture of the triangular flow's generation and the ridge structure's origin in the two-particle correlations.
The approach quantifies the essential feature in the event-by-event fluctuating initial conditions in terms of some local variables instead of a few global deformation parameters, such as eccentricities.
Based on heuristic arguments and numerical simulations, it is argued that the local expansion of a few high-energy spots close to the boundary of the transverse profile primarily generates the triangular flow. 
Each hot spot extends along the longitudinal direction and forms a tube-like structure.
In the framework of the peripheral tube model, the origin of the ridge and shoulder structure can be consistently understood.
In the literature, the long-range two-particle correlations are mainly attributed to higher-order flow harmonics and the essentially linear relation between initial eccentricity and flow anisotropy~\cite{hydro-v3-01, hydro-v3-02, hydro-vn-34, sph-corr-09}.
The peripheral tube model offers a different perspective in this regard.
One might argue that, in terms of the relations between eccentricities and flow harmonics at different orders, it is not straightforward to visualize the underlying physical mechanism.
Specifically, numerical simulations have not convincingly established a well-defined linear response relation for individual orders of azimuthal expansion between the initial conditions and final flow~\cite{sph-vn-04, hydro-vn-34}.
Therefore, the peripheral tube model serves as an alternative to explain the observed features in two-particle correlation.
A more decisive discrimination between the different viewpoints is the evaluation of particle correlations as a function of the emission's location, which is not feasible to date.

The peripheral tube model's initial conditions are constructed reminiscent of those generated by the event generator NEXUS~\cite{nexus-1} and EPOS~\cite{epos-1}.
Specifically, the initial energy distribution is given as a summation of two contributions: the background and peripheral tubes.
In principle, the former can be obtained by averaging the distribution over many events from the same centrality class.
The latter consists of independent high-energy spots close to the system's surface, referred to as the peripheral tubes.
As discussed below, the system's rapid expansion is deflected by the peripheral tube, leading to multiplicity fluctuations and a few flow harmonics associated with the tube.
It is noted that such a phenomenon related to the tube is local, as it will not be perceived by the remainder of the system.
Subsequently, the properties of the entire system are obtained by the superposition of individual tubes. 

Regarding the collective flow, the analytic version of the peripheral tube model~\cite{sph-corr-ev-04} assumes that the hydrodynamic evolution that generates the two-particle correlations is composed of two parts.
The first one is associated with the background energy distribution of the initial conditions, which largely dominates the resultant flow harmonics.
The background particle distribution is mostly governed by an almond shape which generates the elliptic flow primarily.  
On an event-by-event basis, it is also subject to multiplicity fluctuations.
Specifically, the one-particle distribution is not given directly using a Fourier expansion Eq.~\eqref{oneParDis}. 
An individual fluctuating event is given as a sum of two terms: the distribution of the background and that of the tube.
\begin{eqnarray} 
  \frac{dN}{d\varphi}(\varphi,\varphi_\mathrm{tube}) =\frac{dN_{\mathrm{bgd}}}{d\varphi}(\varphi) +\frac{dN_{\mathrm{tube}}}{d\varphi}(\varphi,\varphi_\mathrm{tube}), 
 \label{eq-sec5-1-1}  
\end{eqnarray} 
where 
\begin{eqnarray} 
 \frac{dN_{\mathrm{bgd}}}{d\varphi}(\varphi)&=&\frac{N_\mathrm{bgd}}{2\pi}(1+2v_2^\mathrm{bgd}\cos(2\varphi)),    \label{eq-sec5-1-2}\\  
 \frac{dN_{\mathrm{tube}}}{d\varphi}(\varphi,\varphi_\mathrm{tube})&=&\frac{N_\mathrm{tube}}{2\pi}\sum_{n=2,3}2v_n^\mathrm{tube}\cos(n[\varphi-\varphi_\mathrm{tube}])  . \label{eq-sec5-1-3}
\end{eqnarray} 
In Eq.~\eqref{eq-sec5-1-2}, one considers a simplified scenario where the background flow is essentially the elliptic one, parametrized the flow harmonic $v_2^\mathrm{bgd}$ and the overall multiplicity $N_\mathrm{bgd}$. 
The contributions from the peripheral tube are measured with respect to its angular position $\varphi_\mathrm{tube}$.
To reproduce the desired structure in the two-particle correlations, one introduces a minimal number of Fourier components, encoded in $v_2^\mathrm{tube}$ and $v_3^\mathrm{tube}$. 
It is noted that the contribution associated with the peripheral tube is considered independent between different events, mainly when the event average is performed for the mixed-event correlations. 
We also note that in this model the triangular flow is exclusively generated by the tube. Its symmetry axis is aligned with a fraction of the elliptic flow and both are correlated to the tube's location $\varphi_\mathrm{tube}$.
We assume that the flow from the background is much more significant when compared with those generated by the tube. Therefore, the event plane $\Psi_2$ is primarily governed by the elliptic flow of the background $v_2^\mathrm{bgd}$. 

Following the methods used by the STAR experiment~\cite{RHIC-star-plane-01,RHIC-star-plane-02,RHIC-star-plane-03}, the subtracted di-hadron correlation is defined as
\begin{eqnarray} 
 \langle\frac{dN_{\mathrm{pair}}}{d\Delta\varphi}(\varphi_s)\rangle   =\langle\frac{dN_{\mathrm{pair}}}{d\Delta\varphi}(\varphi_s)\rangle^{\mathrm{prop}} -\langle\frac{dN_{\mathrm{pair}}}{d\Delta\varphi}(\varphi_s)\rangle^{\mathrm{mix}} , \nonumber \\
 \label{eq-sec5-1-4}  
\end{eqnarray} 
where $\varphi_s$ is the trigger particle's angle, which is not integrated out different from most calculations.
By using Eq.(\ref{eq-sec5-1-1}), the proper two-particle correlations can be evaluated by the following form
\begin{eqnarray} 
\langle\frac{dN_{\mathrm{pair}}}{d\Delta\varphi}\rangle^{\mathrm{prop}}  =
 \int\frac{d\varphi_\mathrm{tube}}{2\pi}f(\varphi_\mathrm{tube})
 \frac{dN^T}{d\varphi}(\varphi_s,\varphi_\mathrm{tube})  
\frac{dN^A}{d\varphi} (\varphi_s+\Delta\varphi,\varphi_\mathrm{tube}),   \nonumber \\ \label{eq-sec5-1-corr-proper}
\end{eqnarray}
where $f(\varphi_\mathrm{tube})$ is the distribution function of the tube, and superscripts ``$T$'' and ``$A$'' indicate ``trigger'' and ``associated'' particles respectively.
For instance in a realistic scenario, $f(\varphi_\mathrm{tube})$ might possess a fraction of elliptic modulation owing to the almond-shaped initial condition.

The combinatorial mixed-event two-particle correlations $\langle{dN_{\mathrm{pair}}}/{d\Delta\varphi}\rangle^{\mathrm{mix}}$ can be calculated by using either cumulant or ZYAM method~\cite{zyam-01, zyam-02}. 
While both methods yield very similar results in our model, it is more illustrative to evaluate the cumulant. 

By taking $f(\varphi_\mathrm{tube})=1$ for simplicity, the resultant two-particle correlations read~\cite{sph-corr-ev-04}
\begin{eqnarray}
\langle\frac{dN_\mathrm{pair}}{d\Delta\varphi}(\varphi_s)\rangle^\mathrm{(cmlt)} 
 &=&\frac{\langle N_\mathrm{bgd}^2\rangle-\langle N_\mathrm{bgd}\rangle^2}{(2\pi)^2}(1+2v_2^\mathrm{bgd}\cos(2\varphi_s))
 (1+2v_2^\mathrm{bgd}\cos(2(\Delta\varphi+\varphi_s))) \nonumber\\  
 &+&\left(\frac{N_\mathrm{tube}}{2\pi}\right)^2\sum_{n=2,3}2({v_n^\mathrm{tube}})^2
 \cos(n\Delta\varphi) , 
 \label{ccumulant}
\end{eqnarray} 
where the two terms on the r.h.s. of Eq.~\eqref{ccumulant} are modulated by the background flow harmonics $v_2^\mathrm{bgd}$ and those originated from the tube $v_n^\mathrm{tube}$. 
The first term on the r.h.s. is dominated by the variance of the background multiplicity, namely $\langle N_\mathrm{bgd}^2\rangle -\langle N_\mathrm{bgd}\rangle^2 $.
In other words, the background multiplicity originating from the proper events does not entirely cancel out with those from the mixed event, and the residual is proportional to the overall multiplicity fluctuations.
The first term competes with the second term, which is proportional to $N_\mathrm{tube}$, and measures the event-by-event local fluctuations.
The physical content of the second term exclusively resides on the deflected flow associated with the peripheral tubes.
As the summation starts at $n=2$, the magnitude of $N_\mathrm{tube}$ will not affect the event multiplicity.
Although the second term does not alter the system's overall multiplicity, it modifies the flow harmonics leading to local flow and multiplicity fluctuations.

The above model can be put into the context of the event-plane and centrality dependence of the observed two-particle correlations~\cite{sph-corr-ev-04,sph-corr-ev-06, sph-corr-ev-08}.
In particular, the above expression Eq.~\eqref{ccumulant} explicitly depends on $\varphi_s$, which can be used to study the event plane dependence of the correlation.
The trigonometric dependence of the background contribution on $\varphi_s$ indicates that its contribution to the out-of-plane triggers is opposite to that for the in-plane ones. 
At the out-of-plane direction, there is an overall suppression in the correlations' amplitude, as well as forms a double peak structure on the away side.
Specifically, the observed evolution of the correlation structure on the away side $\Delta\varphi\sim\pi$ can be readily derived as one considers the in-plane trigger with $\varphi_s=0$ and the out-of-plane trigger $\varphi_s=\pi/2$.
Indeed, experimental data~\cite{RHIC-star-plane-01, RHIC-star-plane-02, RHIC-star-plane-03} shows that the overall correlation decreases while the away-side correlation evolves from a single peak to a double peak as $\varphi_s$ increases. 
Since these observed features agree with the analytically derived results, the peripheral model is meaningful despite its simplicity.
We note that such an argument remains mostly unchanged if the ZYAM scheme is employed to subtract the mixed event contribution.

Besides the rational analysis, numerical simulations have also been performed and such a picture has been confirmed~\cite{sph-corr-08, sph-corr-09}.
As shown in Fig.~\ref{fig_tube_evo}, as the system quickly expands it is deflected and generates elliptic and triangular flows that are naturally aligned.

As mentioned above, the flow fluctuations originating from the initial conditions are understood to be of physical relevance.
On the other hand, the overall multiplicity fluctuations are associated mainly with events of different centralities and therefore are less interesting.
In the following section, we elaborate a scheme to normalize the resulting particle correlations by suppressing the multiplicity fluctuations.
Regarding the peripheral tube model, the scheme aims to effectively eliminate the first term on the r.h.s. of Eq.~\eqref{ccumulant}.

\section{A normalization scheme to suppress the multiplicity fluctuations}\label{section3}

In this section, we elaborate on a scheme to suppress the multiplicity fluctuations in calculating two-particle correlations.
The numerical simulations are carried out using the hydrodynamic code SPheRIO~\cite{sph-review-01, sph-review-02}.
These calculations are performed for the initial conditions generated by placing a high-energy tube on top of the background energy-density distribution, reminiscent of the main feature provided by a microscopic event generator, such as NeXuS~\cite{nexus-1,nexus-rept} and EPOS~\cite{epos-1, epos-2}.
Specifically, a fit to the almond shape of the initial conditions for the 20\%-40\% centrality window gives rise to the following energy distribution
\begin{eqnarray}
\epsilon=\epsilon_\mathrm{bgd} + \epsilon_\mathrm{tube}, \label{energyProfile} 
\end{eqnarray}
where
\begin{eqnarray}
\epsilon_\mathrm{bgd} &=& \epsilon_0(9.33+7r^2+2r^4)e^{-r^{1.8}}\ , \label{energyb} 
\end{eqnarray}
with
\begin{eqnarray}
r &=& \sqrt{0.41x^2+0.186y^2} \  .   \nonumber
\end{eqnarray}
and the profile of the high energy tube is calibrated to that of a typical peripheral tube in NeXuS IC
\begin{eqnarray}
\epsilon_\mathrm{tube}&=& 12e^{-(x-x_\mathrm{tube})^2-(y-y_\mathrm{tube})^2} \ , \label{energyt}
\end{eqnarray}
where the tube is located at a given value of energy density close to the surface, whose coordinates on the transverse plane read
\begin{eqnarray}
x_\mathrm{tube} &=& \frac{r_\mathrm{tube}\cos\theta}{\sqrt{0.41\cos^2\theta+0.186\sin^2\theta}}   \\
y_\mathrm{tube} &=& \frac{r_\mathrm{tube}\sin\theta}{\sqrt{0.41\cos^2\theta+0.186\sin^2\theta}} \ . \nonumber
\end{eqnarray}
where the azimuthal angle of the tube $\theta$ is randomized among different events,
and the parameter $r_\mathrm{tube}$ is chosen to be 4.3 fm in the calculations.
Using the above parameterization for the initial conditions, one then solves the hydrodynamic equations with the SPheRIO code and evaluates the resultant two-particle correlations. 
The event-by-event fluctuations of the initial conditions are implemented by introducing a normal distribution in the background
\begin{eqnarray}\label{fluICs}
\epsilon_0 \sim N(\bar{\epsilon}_0 , \sigma^2) ,
\end{eqnarray}
where the mean $\bar{\epsilon}_0 = 1.0$, and the variance $\sigma^2=0.01$, at the initial instant $\tau=1.0$ fm/c.
Here, the position variables, $r, x, y$, are expressed in the unit of fm, and the energy densities $\epsilon$ and $\epsilon_0$ are in the unit of GeV$\cdot$fm$^{-3}$.

The initial conditions are assigned to SPheRIO, where the hydrodynamic equation is solved numerically on an event-by-event basis and the fluctuations in the initial energy distribution are considered.
We generate 1000 events to simulate the 200 A GeV Au-Au collisions. 
At the end of the hydrodynamic evolution of each event, a Monte-Carlo generator is employed to perform hadron emission in the Cooper-Frye prescription. 
In our calculations, the Monte-Carlo process is carried out 50 times for a given event.
Subsequently, the hadron decay is also considered. 

\begin{figure}[ht]
\begin{tabular}{cc}
\vspace{-26pt}
\begin{minipage}{250pt}
\centerline{\includegraphics[width=1.2\textwidth,height=1.0\textwidth]{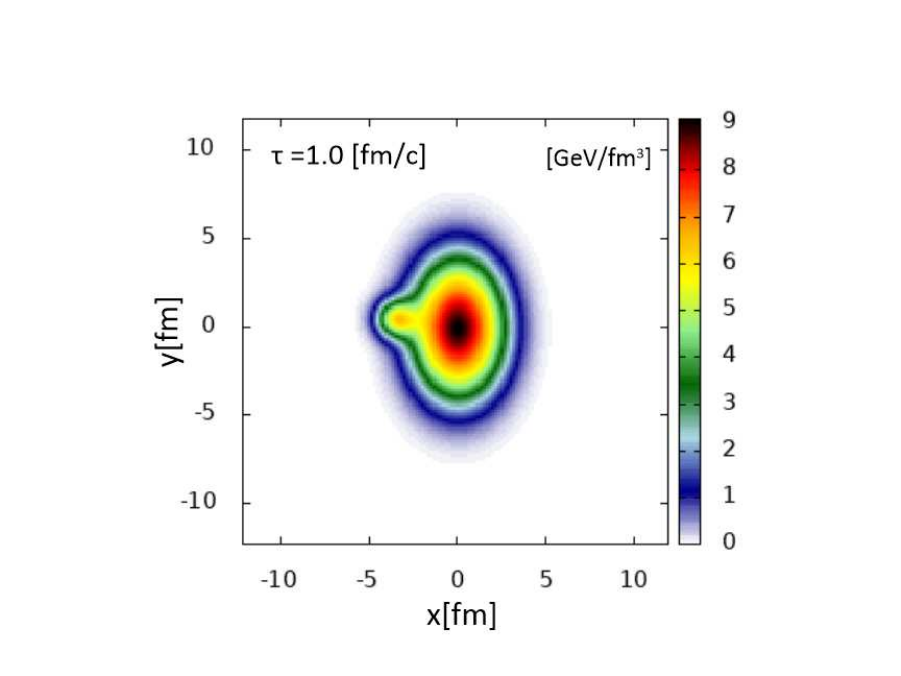}}
\end{minipage}
&
\vspace{26pt}
\begin{minipage}{250pt}
\centerline{\includegraphics[width=1.2\textwidth,height=1.0\textwidth]{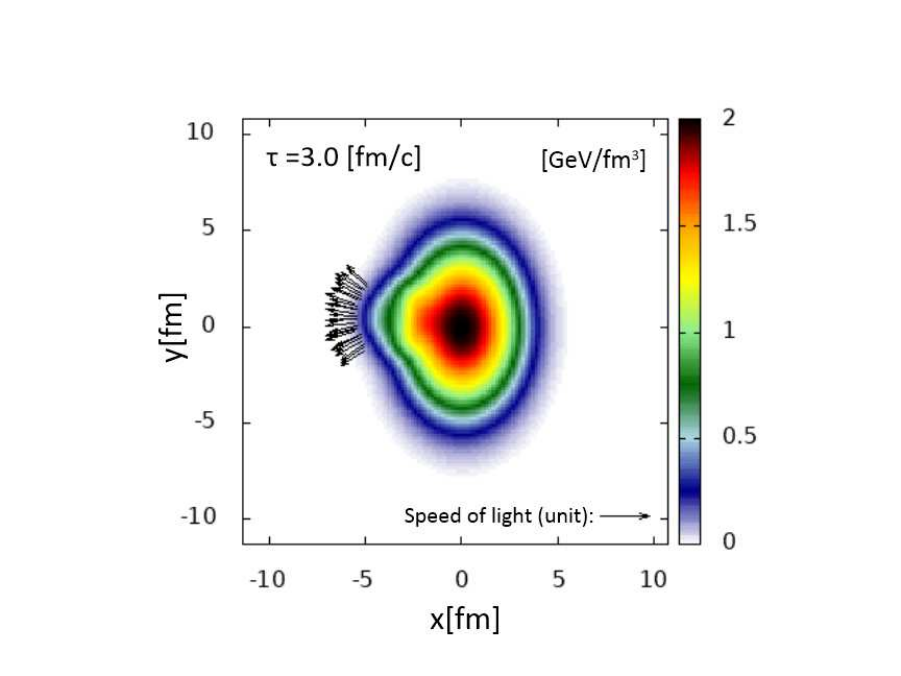}}
\end{minipage}\\
\begin{minipage}{250pt}
\centerline{\includegraphics[width=1.2\textwidth,height=1.0\textwidth]{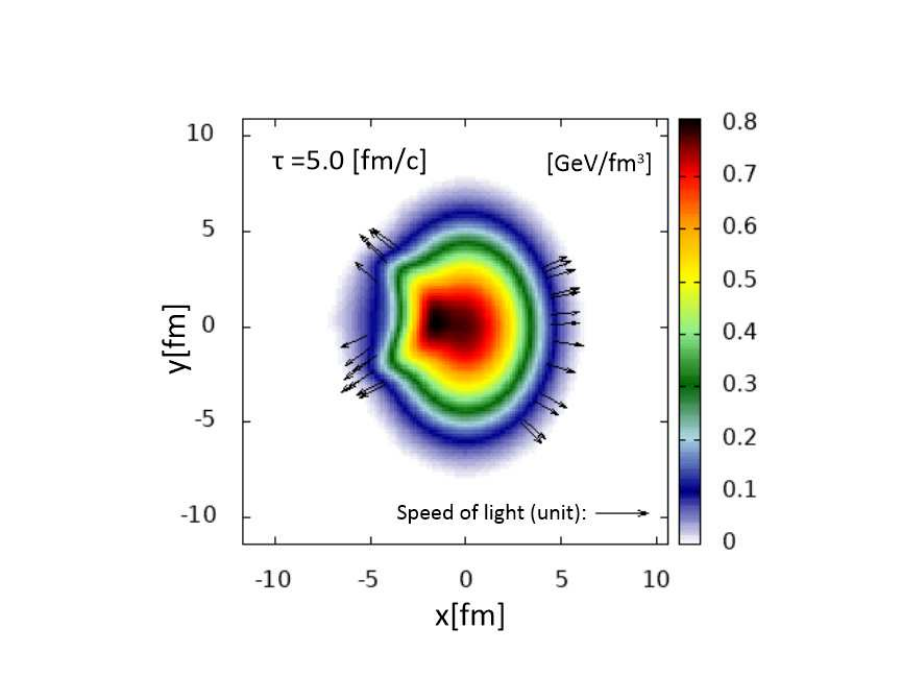}}
\end{minipage}
&
\vspace{26pt}
\begin{minipage}{250pt}
\centerline{\includegraphics[width=1.2\textwidth,height=1.0\textwidth]{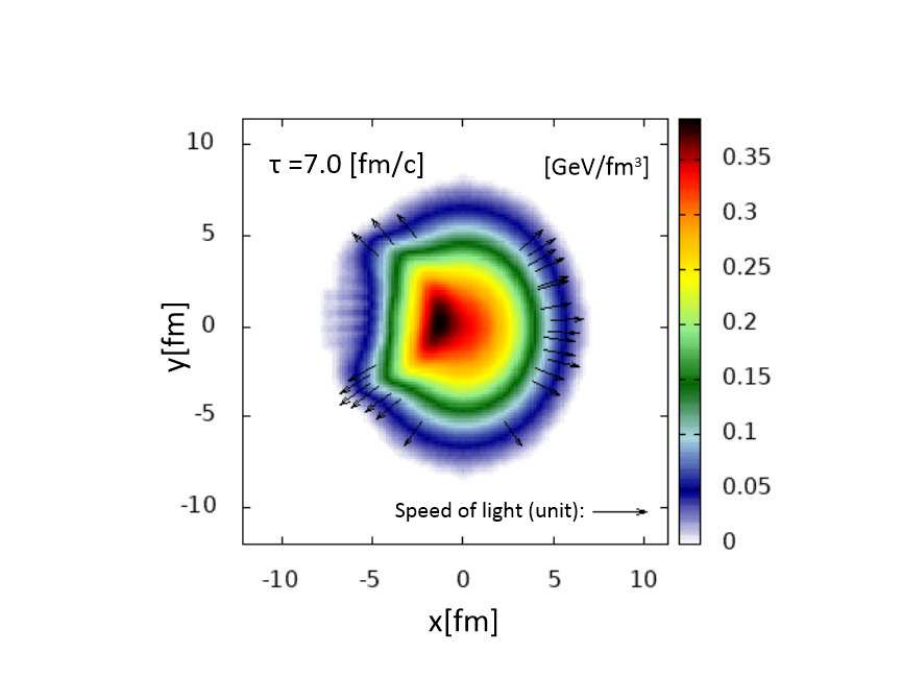}}
\end{minipage}
\end{tabular}
\renewcommand{\figurename}{Fig.}
\caption{The peripheral tube model's initial conditions and the temporal evolutions evaluated at different instants $\tau = 1.0, 3.0, 5.0$, and $7.0$ fm/c.
The color scale describes the energy density expressed in GeV$\cdot$fm$^{-3}$.
The vectors represent the fluid velocity, whose magnitude is shown relative to that of the speed of light, indicated by the vector at the bottom-right corner.}
\label{fig_tube_evo}
\end{figure}

In Fig.~\ref{fig_tube_evo}, we present the peripheral tube model's initial conditions and the temporal evolutions at different time instants. While the system expands one observes that the flow is deflected by the high-energy tube, which gives rise to a local triangular component.
The velocity vectors indicate that the flow near the tube is primarily directed in two directions forming an angle of approximately $60$ degrees, observed in the profiles at $\tau=5.0$ and $7.0$ fm/c. 
As discussed below, the latter is also demonstrated in the resultant two-particle correlations.

To evaluate di-hadron correlations, the combinatorial background is evaluated using the two-particle cumulant. 
The background modulation is then subtracted from the proper two-particle correlations according to Eq.~\eqref{eq-sec5-1-1}.

In Fig.~\ref{fig_p2_fluc_supp}, we present the subtracted two-particle correlations in solid black curves.
The obtained correlation structures are evaluated and shown for different trigger angles from the in-plane to the out-of-plane directions.
The calculated correlations as a function of $\Delta\varphi$ vary for different trigger directions, particularly on the away side.
In the in-plane direction, the correlations exhibit a single peak on the away side that is broader than the near-side one.
The correlation structure on the away side continuously involves forming a double peak in the out-of-plane direction.
This characteristic is in agreement with the STAR measurements.

However, one may also perform the numerical simulations by manually removing the event-by-event fluctuations in the background.
This can be achieved by assigning $\sigma=0$ in Eq.~\eqref{fluICs}.
The rest of the calculations essentially remain the same.
The red dashed curves in Fig.~\ref{fig_p2_fluc_supp} show the resultant subtracted two-particle correlations.

When compared against the solid black curves, it is observed that the evolution of the correlation structure from the in-plane to the out-of-plane directions is significantly suppressed.
As a matter of fact, the two-particle correlations mostly remain unchanged.
This can be intuitively understood by observing Eq.~\eqref{ccumulant}.
The first term on the r.h.s. of Eq.~\eqref{ccumulant} is proportional to the background multiplicity fluctuations, primarily governed by the fluctuations Eq.~\eqref{fluICs} in the initial conditions.
The second term on the r.h.s. of Eq.~\eqref{ccumulant} is governed mainly by the local flow deflection associated with the tube, which stays unchanged for different events.
As one removes the overall fluctuations by taking $\sigma=0$ in Eq.~\eqref{fluICs}, the first term on the r.h.s. of Eq.~\eqref{ccumulant} is significantly suppressed.
As a result, the two-particle correlations are less significant in magnitude and essentially governed by elliptic and triangular flow components related to the tube.
In other words, the numerical simulations indicate that analysis leading to Eq.~\eqref{ccumulant} is relevant.

\begin{figure}[ht]
\begin{minipage}{210pt}
\centerline{\includegraphics[height=1.2\textwidth]{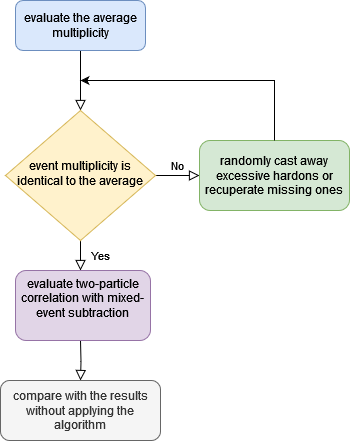}}
\end{minipage}
\renewcommand{\figurename}{Fig.}
\caption{The flowchart of the proposed suppression algorithm.}
\label{fig_flowchart}
\end{figure}

\begin{figure}[ht]
\begin{tabular}{cc}
\vspace{-22pt}
\begin{minipage}{210pt}
\centerline{\includegraphics[width=1.2\textwidth,height=1.0\textwidth]{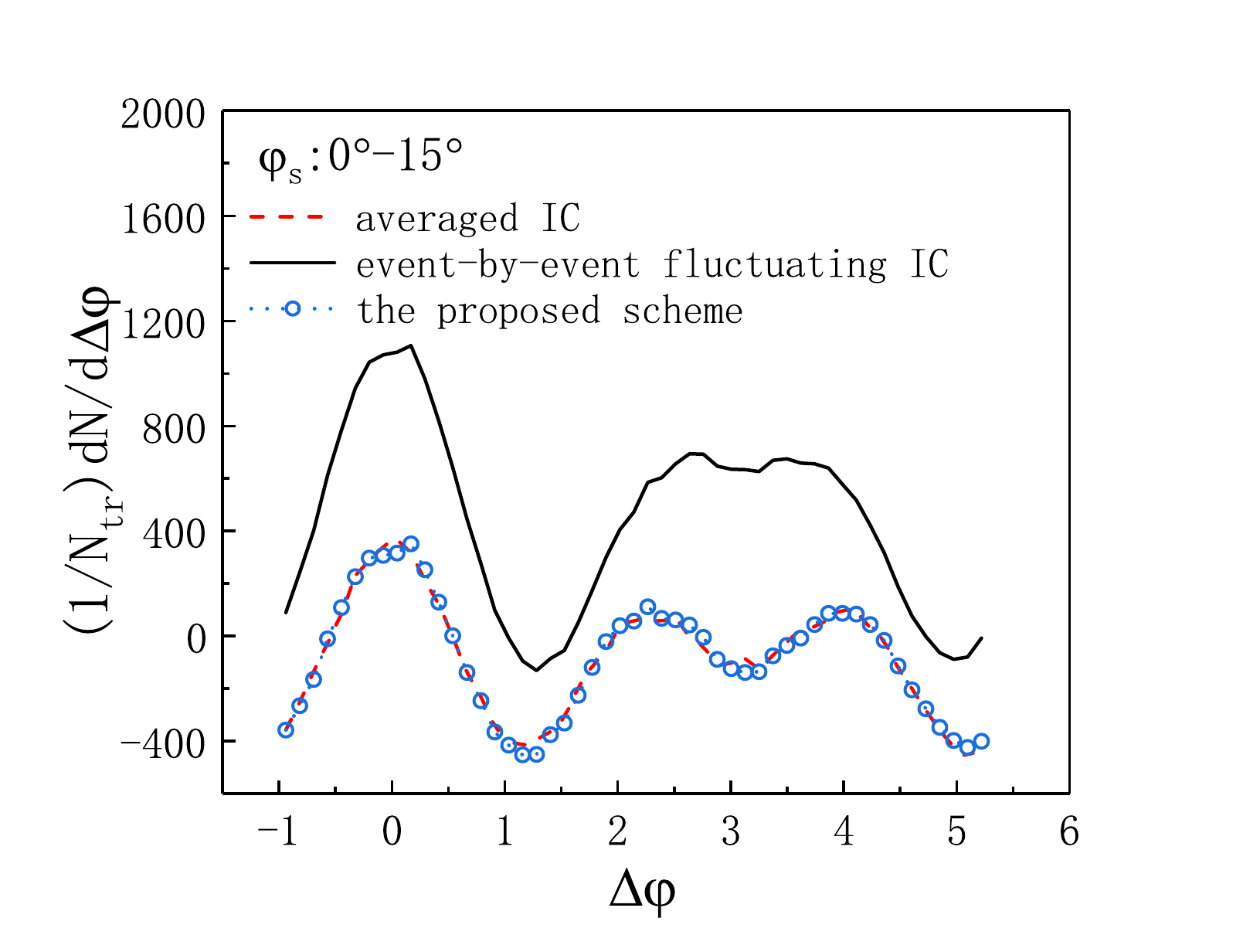}}
\end{minipage}
&
\vspace{15pt}
\begin{minipage}{210pt}
\centerline{\includegraphics[width=1.2\textwidth,height=1.0\textwidth]{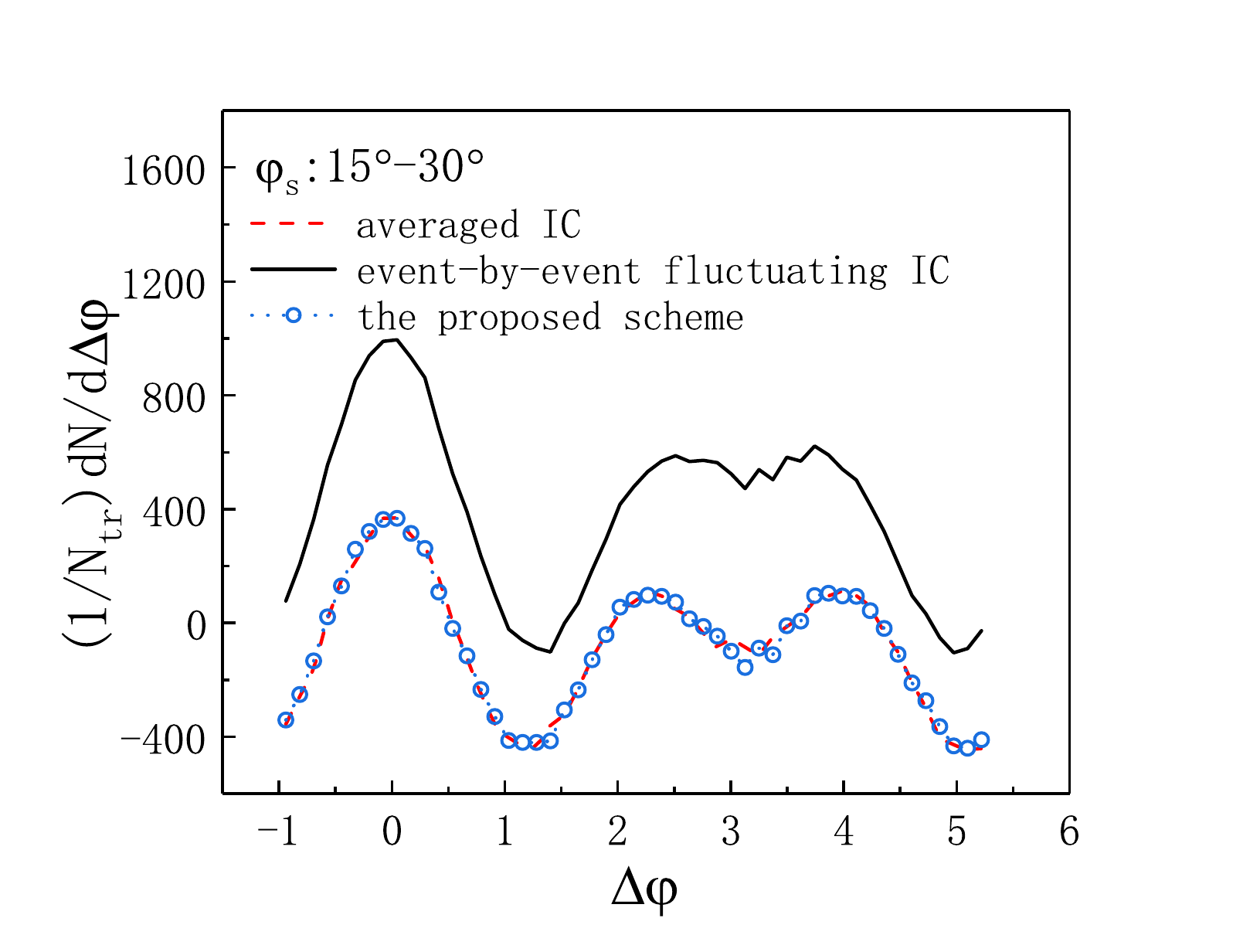}}
\end{minipage}
\\
\vspace{15pt}
\begin{minipage}{210pt}
\centerline{\includegraphics[width=1.2\textwidth,height=1.0\textwidth]{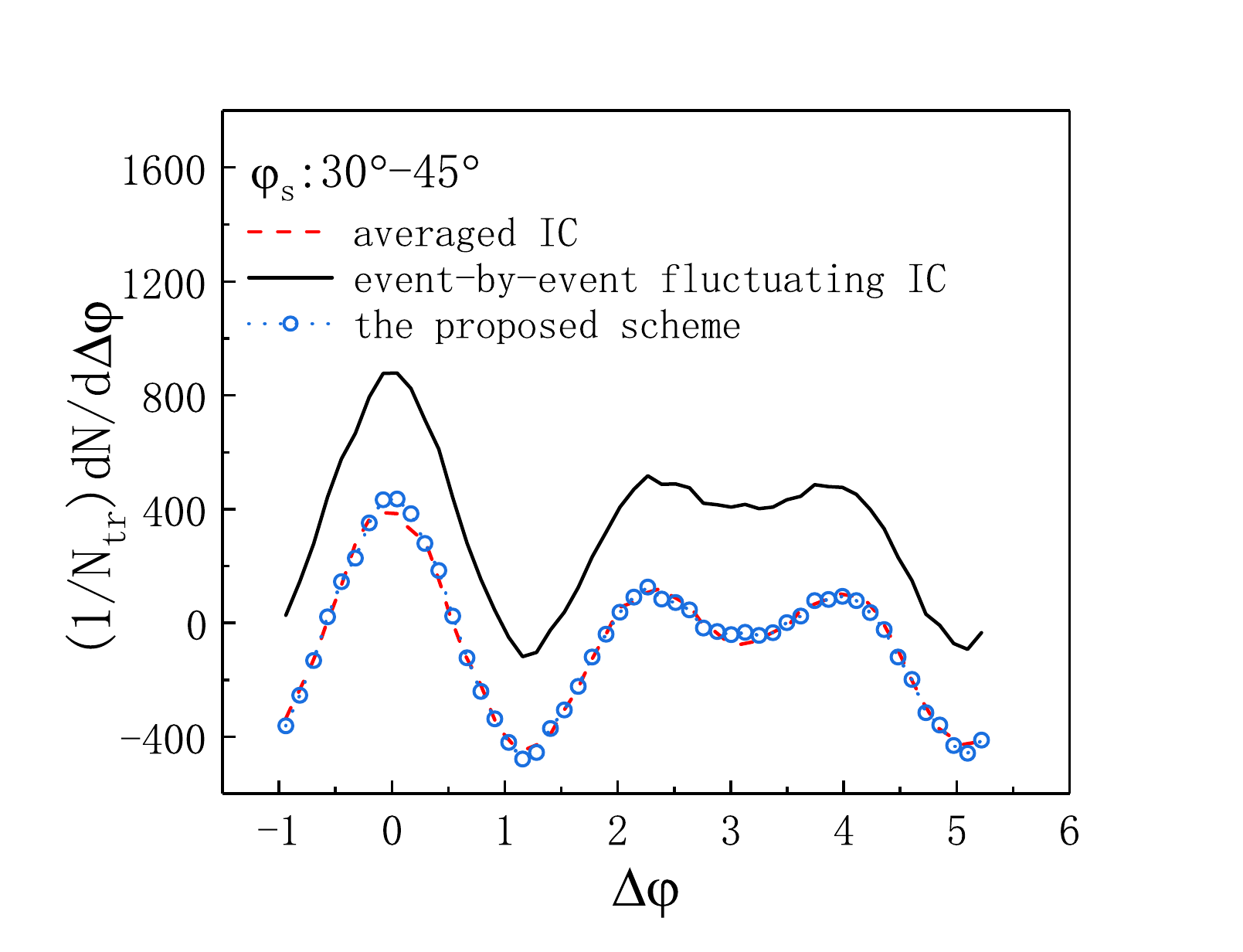}}
\end{minipage}
&
\begin{minipage}{210pt}
\centerline{\includegraphics[width=1.2\textwidth,height=1.0\textwidth]{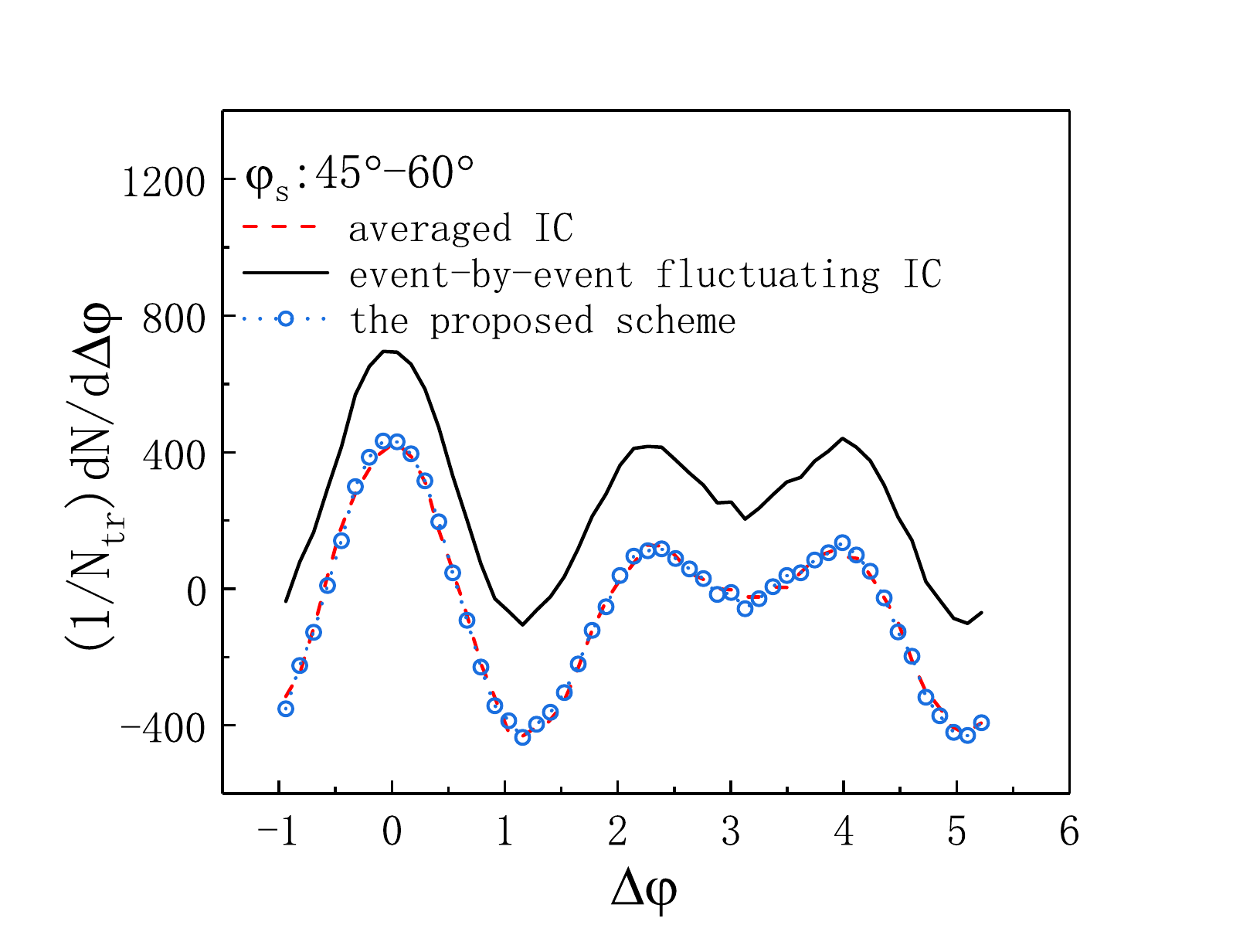}}
\end{minipage}
\\
\vspace{15pt}
\begin{minipage}{210pt}
\centerline{\includegraphics[width=1.2\textwidth,height=1.0\textwidth]{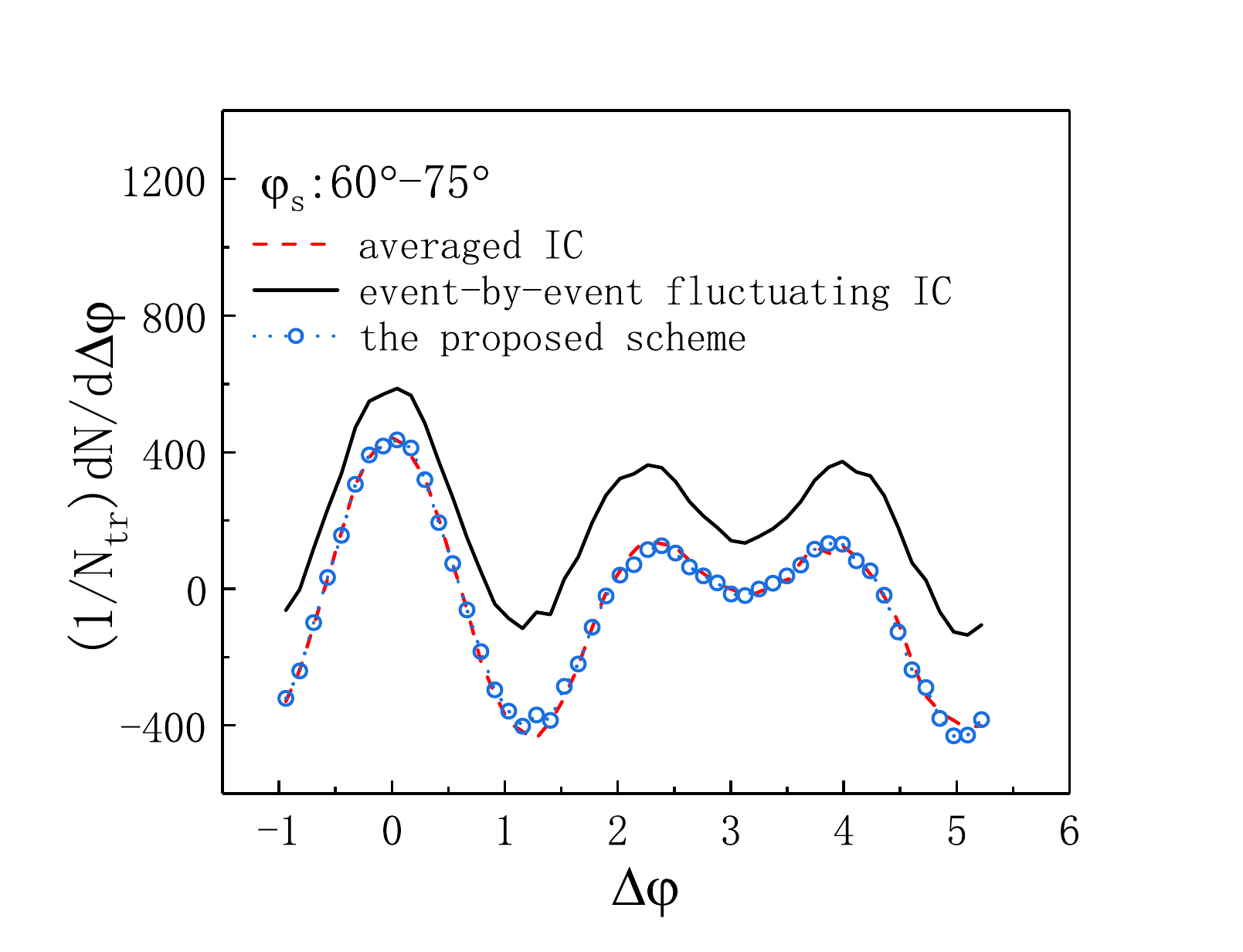}}
\end{minipage}
&
\vspace{15pt}
\begin{minipage}{210pt}
\centerline{\includegraphics[width=1.2\textwidth,height=1.0\textwidth]{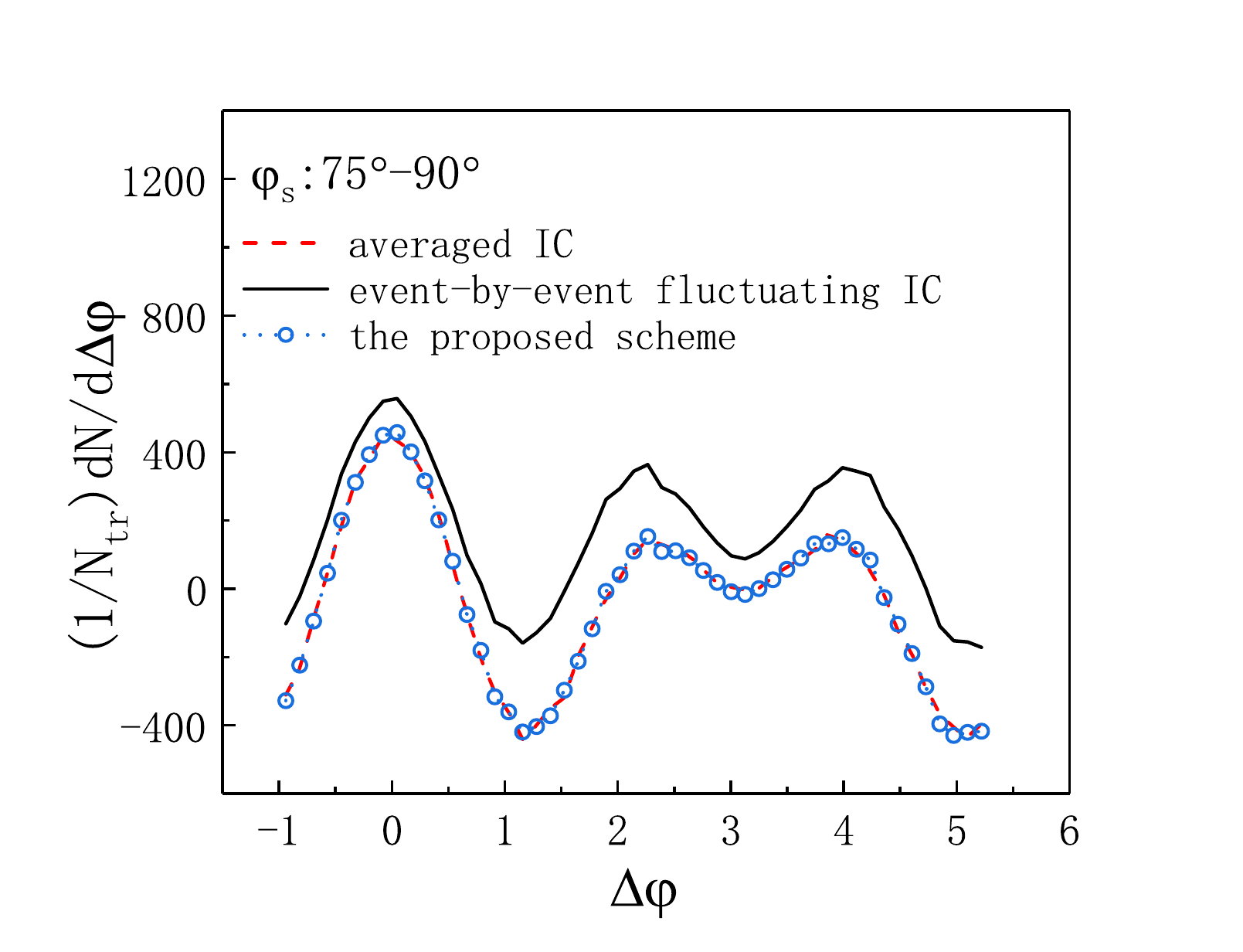}}
\end{minipage}
\end{tabular}
\renewcommand{\figurename}{Fig.}
\caption{The subtracted two-particle correlations for three different scenarios.
The two-particle cumulants calculated for the event-by-event fluctuating initial conditions (denoted as IC in the legend) are shown in solid black curves.
Those obtained for averaged initial conditions are presented in dashed red curves.
The results of implementing the proposed scheme for the event-by-event fluctuating initial conditions are shown in empty blue circles.}
\label{fig_p2_fluc_supp}
\end{figure}

Based on the above arguments, we propose the following scheme for calculating the two-particle correlations using cumulant as shown in Fig.~\ref{fig_flowchart}.
Instead of using all the emitted hadrons, one normalizes the multiplicity of different events to their average value when evaluating particle pairs.
Specifically, if one event has more hadrons than the average number, excessive ones will be cast away.
These hadrons will be chosen randomly.
On the other hand, if one event has fewer hadrons than the average number, the insufficiency will be compensated by randomly choosing some hadrons for a second time. 
By employing the above strategy, one always has the same number of hadron pairs in calculating the cumulant.
The multiplicity fluctuations will be effectively removed.

In Fig.~\ref{fig_p2_fluc_supp}, the empty blue circles indicate the resultant two-particle correlations when the proposed suppression scheme is implemented.
The results are satisfactory, as the empty blue circles essentially sit on top of the red dashed curves.
The subtracted two-particle correlations do not vary significantly as one goes from the in-plane to the out-of-plane direction.
Compared to the analytic result, it is understood to be essentially governed by the r.h.s. of Eq.~\eqref{ccumulant}.

Before closing this section, we note that there is an alternative perspective to understand the above normalization process.
In general, the multi-particle correlators and cumulants can be analyzed in terms of the moments~\cite{hydro-corr-ph-23}
\begin{eqnarray}\label{NparCorrMon}
\mu_{n_1,n_2,\cdots,n_N} = \langle e^{i\left(n_1\varphi_1+n_2\varphi_2+\cdots+n_N\varphi_N\right)}\rangle ,
\end{eqnarray}
which can be further evaluated using Q-vectors~\cite{Bilandzic:2013kga}
\begin{eqnarray}\label{defQV}
Q_{n, p} = \sum_{k=1}^M w_k^p e^{in\varphi_k} .
\end{eqnarray}
Similar to the normalization in the two-particle correlations, the moments defined in Q-vectors must be calculated in a multiplicity-independent fashion.
For instance, the calculation involves the following term (c.f. Eq.~(19) of~\cite{Bilandzic:2013kga})
\begin{eqnarray}\label{defQVN}
N\langle 2\rangle_{n_1, n_2} = Q_{n_1, 1}Q_{n_2, 1}-Q_{n_1+n_2, 2}=\sum_{k_1=1}^M e^{in_1\varphi_{k_1}}\sum_{k_2=1}^M e^{in_2\varphi_{k_2}}-\sum_{k_3=1}^M e^{i(n_1+n_2)\varphi_{k_3}} ,
\end{eqnarray}
must be normalized by the denominator
\begin{eqnarray}\label{defQVD}
D\langle 2\rangle_{n_1, n_2} = N\langle 2\rangle_{0, 0} ,
\end{eqnarray}
so that multiplicity does not enter the ratio $N\langle 2\rangle_{n_1, n_2}/D\langle 2\rangle_{n_1, n_2}$.

Also, the event-average part of the cumulant does not explicitly involve multiplicity and its fluctuations.
As an example, the two-particle cumulant is defined as follows~\cite{hydro-corr-ph-04}
\begin{eqnarray}\label{defCumulant}
\langle\langle e^{in(\varphi_1-\varphi_2)}\rangle\rangle = \langle e^{in(\varphi_1-\varphi_2)}\rangle -\langle e^{in \varphi_1}\rangle\langle e^{-in \varphi_2}\rangle ,
\end{eqnarray}
where the average $\langle\cdots\rangle$ includes the average involving all the hadrons for a given event and the event average.
For the latter, the difference in multiplicity between different events should not play a role.

\section{Concluding remarks}\label{section4}

In this study, we elaborate on a scheme to suppress the multiplicity fluctuations in the two-particle correlations.
The suppression is achieved by adequately normalizing the multiplicity when forming the pairs for the proper and mixed events in the calculations of particle correlations.
As a result, the cumulants are obtained in a manner irrelevant to multiplicity.
As the multiplicity fluctuations are adequately extracted, the resulting two-particle correlations demonstrate the main features governed by the peripheral tube model.
The latter was shown numerically by using Monte Carlo simulations.
Such a normalization scheme naturally agrees with the multi-particle correlators typically evaluated in terms of the ratio of the Q-vectors.

The present paper is mainly devoted to analyzing particle correlations in the context where the collective flow dominates.
Although essentially indistinguishable from an empirical perspective, particle correlations emerge due to different physical origins.
In relativistic heavy-ion collisions, collective flow, conservation laws, relics of minijets, and HBT interferometry all contribute to the resultant two-particle correlations, particularly the ``ridge''.
To effectively obtain the relevant physical content, it is essential to develop observables for which specific irrelevant contributions to the fluctuations can be suppressed or separated.
In this regard, one can utilize the empirical observables to extract information on the initial states, which also feature intrinsic fluctuations owing to the underlying microscopic physical system.
The attempt to separate the contributions of different physical mechanisms has been a continuous effort from both the theoretical and experimental sides.
The role of collective flow on particle correlation and the related cumulant approach was first elaborated and promoted by Borghini, Ollitrault {\it et al.}~\cite{hydro-corr-ph-03, hydro-corr-ph-04}.
It was proposed to use particle correlation to extract the information on flow harmonics and was further developed~\cite{Bilandzic:2010jr} regarding the use of Q-vectors besides the generating functional in practice.
These studies were primarily based on the assumption that the collective flow entirely governs the observed particle correlations.
As pointed out in Ref.~\cite{hydro-corr-ph-03}, the non-flow's impact typically becomes suppressed by the flow as the event multiplicity increases and as one considers multi-particle correlations.
Experimentalists have widely used these cumulant approaches to obtain flow harmonics involving an average between events of different multiplicities, where multiplicity fluctuations inevitably play a role.
In Ref.~\cite{Bilandzic:2010jr}, it was suggested that one might assign a weight to the event average, while the best approach might be to calculate the cumulants at fixed multiplicity and then average over the entire event sample.
For a given centrality window, this is because that the multiplicity fluctuations are mainly of statistical origin.
The latter can be attributed to the finite size of the centrality window or the finite multiplicity~\cite{sph-bes-01, sph-bes-02, sph-vn-09, sph-vn-10}, which are largely irrelevant to the physics of interest.
In this regard, the proposed method is inspired by the above observation and the analytic results encountered in~\cite{sph-corr-ev-04}.

In our calculations, we only elaborate on two-particle cumulant.
But following this train of thought, one may generalize the two-particle cumulant
\begin{equation}
\llangle e^{i2(\varphi_1-\varphi_2)}\rrangle \nonumber
\end{equation} 
to the four-particle cumulant 
\begin{equation}
\llangle e^{i2(\varphi_1+\varphi_2-\varphi_3-\varphi_4)}\rrangle = \langle e^{i2(\varphi_1+\varphi_2-\varphi_3-\varphi_4)}\rangle - \llangle e^{i2(\varphi_1-\varphi_3)}\rrangle \llangle e^{i2(\varphi_2-\varphi_4)}\rrangle - \llangle e^{i2(\varphi_1-\varphi_4)}\rrangle \llangle e^{i2(\varphi_2-\varphi_3)}\rrangle , \nonumber
\end{equation} 
using the same reasoning that was employed to derive Eq.~\eqref{ccumulant}. 
Again, mixed event correlation will not include any contribution between the background and the tube, as it cancels out in the event-by-event average.
Intuitively, the resulting expression will involve higher moments of the multiplicity fluctuations such as $\langle N_\mathrm{bgd}^4\rangle-\langle N_\mathrm{bgd}^2\rangle^2$, which is expressed in terms of angle differences $\varphi_1-\varphi_4$, $\varphi_2-\varphi_3$, and $\varphi_1-\varphi_3$.
Unfortunately, to our knowledge, it seems that such measurements are not straightforward and have not been performed experimentally.

Regarding the application of the proposed method, the present study involves a scheme primarily devoted to suppressing the multiplicity fluctuations in A+A collisions.
For such a scenario, events are typically characterized by high multiplicity, ensuring that collective flow emerges and dominates the resultant particle correlations.
Moreover, the numerical simulations presented in this work show that the proposed scheme can indeed effectively suppress the multiplicity and extract relevant information on particle correlations.
The good agreement with the theoretical results is not difficult to understand because the latter's dynamics are primarily hydrodynamics.
It is also noted that one can always normalize residual correlations by the number of particle pairs to compare to the simulations or experimental data.
On the other hand, one must point out that one encounters a somewhat different physical picture for small systems.
The magnitude of the multiplicity of the underlying event may determine whether collectivity plays a role in the produced matter.
Events with different multiplicities are likely to be governed by entirely distinct physical mechanisms.
Even in the framework of the tube model, if one randomizes the number of tubes to the extent that it affects the overall initial geometric shape of the underlying event, the multiplicity fluctuations will be entangled with those of flow harmonics and substantially affecting the validity of the approach.
Therefore, special care should be taken regarding the apparent suppression of multiplicity fluctuations between such events by employing the proposed algorithm, which might indiscriminately eliminate crucial physical content.
Nonetheless, we understand that the proposed approach can be applied to the same extent as cumulant method.
Even with a reasonable degree of flow fluctuations, the procedure can be performed to extract flow harmonics, as the experimentalist does.
In this context, the method can be viewed as an alternative to the existing approaches. Specifically, when comparing its deviation from the others, a measure of uncertainty between available means to extract flow harmonics from the data.
Regarding numerical simulations, it is relevant to perform more detailed analyses regarding different profiles of initial fluctuations, the dependence of the resulting correlations on transverse momentum cuts, and possible comparisons between models of hydrodynamic and transport origins.
We plan to carry out such studies in future works.

\section*{Acknowledgements}

We extend our sincere gratitude to Ted William Grant for his meticulous proofreading of the manuscript.
We gratefully acknowledge the financial support from Brazilian agencies 
Funda\c{c}\~ao de Amparo \`a Pesquisa do Estado de S\~ao Paulo (FAPESP), 
Funda\c{c}\~ao de Amparo \`a Pesquisa do Estado do Rio de Janeiro (FAPERJ), 
Conselho Nacional de Desenvolvimento Cient\'{\i}fico e Tecnol\'ogico (CNPq), 
and Coordena\c{c}\~ao de Aperfei\c{c}oamento de Pessoal de N\'ivel Superior (CAPES).
This work is also supported by the National Natural Science Foundation of China  (Grants Nos. 12047506 and 12205032).
This work is partially supported by the Natural Science Foundation of Guangxi
(No. 2020GXNSFBA159003) and the Central Government Guidance Funds for Local Scientific and Technological Development, China (No. Guike ZY22096024).
A part of this work was developed under the project Institutos Nacionais de Ciências e Tecnologia - Física Nuclear e Aplicações (INCT/FNA) Proc. No. 464898/2014-5.
This research is also supported by the Center for Scientific Computing (NCC/GridUNESP) of S\~ao Paulo State University (UNESP).
This work is also supported by the Postgraduate Research \& Practice Innovation Program of Jiangsu Province under Grant No. KYCX22-3453.


\end{document}